# Pole Positions of the Δ(1232) Resonance

Mohamed E. Kelabi[1]

**Abstract**
The phase shift of pion nucleon elastic scattering $\delta_{33}$, corresponding to isospin $I=3/2$ and angular momentum $J=3/2$, has been parameterized over the range $1100 < W < 1375$ MeV, using the single channel $p\pi^+$ data from CNS-DAC. By employing our four parameters formula, the pole positions of corresponding to the Δ(1232) resonance are then identified.

## 1.1 Introduction

In the elastic scattering, the change of amplitudes and phases of the outgoing waves, is commonly expressed by the relation [1], [2]

$$\eta_l = e^{2i\delta_l} = 1 + 2i \sin\delta_l \, e^{i\delta_l}. \qquad (1.1)$$

where $\delta_l$ is the $l^{\text{th}}$ phase shift. Insert Eq. (1.1) into the scattering amplitude,

$$f_l = \frac{\eta_l - 1}{2iq},$$

gives

$$f_l = \frac{\sin\delta_l \, e^{i\delta_l}}{q} \qquad (1.2)$$

where $q$ is the momentum of pion in the CM frame.

---

[1] Physics Department, Al-Fateh University, Tripoli, LIBYA.

## 1.2 Formalism and Results

The inverse of the scattering amplitude Eq. (1.2) can be rewritten as

$$\frac{1}{f_l} = K_l^{-1} - iq \tag{1.3}$$

where we have introduced

$$K_l^{-1} = q \cot \delta_l, \tag{1.4}$$

which is an analytic function of $W^2$ and hence $q^2$ at threshold. To investigate the energy dependence of the resonant amplitude, we define

$$H(W) = q^{2l+1} \cot \delta_l = q^{2l} K_l^{-1}. \tag{1.5}$$

This is important for resonances near the threshold, since that provided $H(W)$ is finite at threshold and, it will automatically give the correct threshold behaviour for $K_l^{-1}$, and so for the scattering amplitude[2]. To locate the pole position properly, we constrict on accurate parameterization of the phase shift data. Specifically we consider the $p$-wave parameterization:

$$q^3 \cot \delta = a_0 + a_2 q^2 + a_4 q^4 + a_6 q^6 \tag{1.6}$$

where the $\pi N$ phase shift $\delta \equiv \delta_{2I\,2J} = \delta_{33}$, as measured in $p\pi^+$ scattering. The result of parameterization gives

$$\begin{aligned}
a_0 &= \phantom{-}5.19909 \pm 0.05182 \\
a_2 &= -0.78451 \pm 0.12418 \\
a_4 &= -0.31100 \pm 0.06212 \\
a_6 &= -0.05150 \pm 0.00818
\end{aligned} \tag{1.7}$$

in charged pion mass units. The fit is based on the chi-squared per degrees of freedom,

$$\chi^2 \equiv \frac{1}{m-p} \sum_m \left( \frac{exp_m - fit_m}{\varepsilon_m} \right)^2$$

where $m$, $p$, and $\varepsilon$ are the number of data points, degrees of freedom, and the corresponding error on each data point, respectively. The fit gives $\chi^2 = 2$. In Figure 1-1, we plot our parameterization of the $\delta_{33}$ compared with $p\pi^+$ single channel data[3].

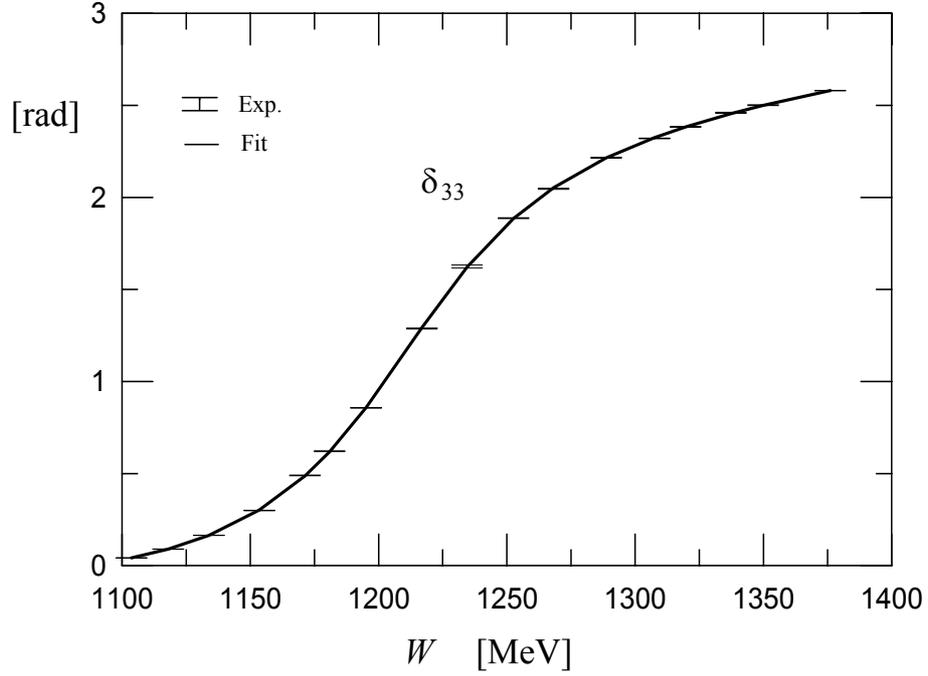

Figure 1-1. Phase shift parameterization of the $\delta_{33}$. The value of error on each data point is presented by the symbol ⊢ .

To locate the pole position, we seek the zeros of reciprocal of the scattering amplitude,

$$\frac{1}{f_{33}} = q^3 \cot\delta_{33} - iq = 0, \qquad (1.8)$$

at the poles. Substituting Eq. (1.6) into Eq. (1.8), gives

$$a_0 + a_2 q'^2 + a_4 q'^4 + a_6 q'^6 - iq'^3 = 0 \qquad (1.9)$$

as the equation determining the position $q'$ of any poles. Inserting the parameters obtained Eq. (1.7) into Eq. (1.9) and solving for $q'$ of the form

$$q' = a + ib \qquad (1.10)$$

gives

$$q' = \begin{cases} \pm 1.69822 - i\, 2.13093 \\ \pm 1.50154 - i\, 0.34376 \\ \phantom{\pm} i\, 3.10238 \\ \phantom{\pm} i\, 1.84699. \end{cases} \qquad (1.11)$$

The corresponding values of the total CM energy

$$W = \sqrt{q'^2 + m_N^2} + \sqrt{q'^2 + m_\pi^2}$$

can be obtained using the $\pi^+ p$ kinematics. These are given below in the units of [MeV]:

$$W = \begin{cases} 1177.81 \pm i\,354.10 \\ 1210.67 \pm i\,50.62 \\ 832.38 - i\,409.89 \\ 902.16 + i\,216.73. \end{cases} \quad (1.12)$$

Only the upper middle values are close to the Δ-resonance position "$W \approx 1232$ MeV", we therefore identify that

$$W_r = 1210.67 \text{ MeV}, \quad \Gamma_r/2 = 50.62 \text{ MeV} \quad (1.13)$$

as the resonance pole positions. These values are comparable with other existing fits, shown in Table 1-1.

| Real Part | − 2 × Imaginary Part | Name | Fit |
|---|---|---|---|
| 1210.67 ± 0.42 | 101.2 ± 1.0 | Present work | 2006 |
| 1211±1 to 1212±1 | 102±2 to 99±2 | Hanstein[4] | 1996 |
| 1206.9±0.9 to 1210.5±1.8 | 111.2±2.0 to 116.6±2.2 | Miroshnichenko[5] | 1979 |
| 1208.0±2.0 | 106±4 | Campbell[6] | 1976 |

Table 1-1. Pole positions comparison for $\Delta^+$ (1232).

**1.3 Conclusion**

The given parameterization shows a good fit compared with the experiment of $p\pi^+$, over the available energy range $1100 < W < 1375$ MeV. The obtained phase shift $\delta_{33}$ can also be employed, using other appropriate techniques, to extract the position and width of the Δ(1232) resonance. Further improvement of the fit may be achieved by adding more terms.

**Acknowledgement**
I am grateful to Dr. Graham Shaw[2] for his valuable comments.

---

[2] Department of physics and Astronomy, University of Manchester, Manchester, UK.